\documentclass[conference]{IEEEtran}
\IEEEoverridecommandlockouts
\usepackage{cite}
\usepackage{amsmath,amssymb,amsfonts}
\usepackage{algorithmic}
\usepackage{graphicx}
\usepackage{textcomp}
\usepackage{xcolor}

\usepackage{ctable}
\usepackage{multirow}

\newcommand\norm[1]{\left\lVert#1\right\rVert}

\def\BibTeX{{\rm B\kern-.05em{\sc i\kern-.025em b}\kern-.08em
    T\kern-.1667em\lower.7ex\hbox{E}\kern-.125emX}}
\begin{document}

\title{Cosine-similarity penalty to discriminate sound classes in weakly-supervised sound event detection
}

\author{\IEEEauthorblockN{Thomas Pellegrini}
\IEEEauthorblockA{\textit{IRIT, Universit\'e Paul Sabatier, CNRS}\\
Toulouse, France \\
thomas.pellegrini@irit.fr}
\and
\IEEEauthorblockN{L\'eo Cances}
\IEEEauthorblockA{\textit{IRIT, Universit\'e Paul Sabatier, CNRS}\\
Toulouse, France \\
leo.cances@irit.fr}

}

\maketitle

\begin{abstract}
The design of new methods and models when only weakly-labeled data are available is of paramount importance in order to reduce the costs of manual annotation and the considerable human effort associated with it. In this work, we address Sound Event Detection in the case where a weakly annotated dataset is available for training. The weak annotations provide tags of audio events but do not provide temporal boundaries. The objective is twofold: 1) audio tagging, i.e. multi-label classification at recording level, 2) sound event detection, i.e. localization of the event boundaries within the recordings. This work  focuses mainly on the second objective. We explore an approach inspired by Multiple Instance Learning, in which we train a convolutional recurrent neural network to give predictions at frame-level, using a custom loss function based on the weak labels and the statistics of the frame-based predictions. Since some sound classes cannot be distinguished with this approach, we improve the method by penalizing similarity between the predictions of the positive classes during training. On the test set used in the DCASE 2018 challenge, consisting of 288 recordings and 10 sound classes, the addition of a penalty resulted in a localization F-score 
of 34.75\%, and brought 10\% relative improvement compared to not using the penalty. Our best model achieved a 26.20\% F-score on the DCASE-2018 official Eval subset close to the 10-system ensemble approach that ranked second in the challenge with a 29.9\% F-score. 
\end{abstract}



\section{Introduction}
\label{sec:intro}

With today's technologies, large corpora of annotated audio data are still required to train state of the art models such as deep neural networks (DNN). This holds true in many audio related applications, e.g. speech recognition, acoustic scene classification and sound event detection (SED), the topic covered in this article. In speech processing, a bunch of research work has been devoted to design automatic tools to process the so-called under-resourced or less-represented languages~\cite{pellegrini2009automatic,besacier2014automatic}. In the context of SED, \textit{low resource} may refer to different situations: a lack of audio data, a lack of reference annotations created manually (\textit{ground truth labels}), the availability of partial annotations only. In this article, we are concerned with this last situation where so-called \textit{weak labels} only are available. In SED, weak labels refer to ground truth labels that describe a recording at a global level, with no temporal information on the onset nor offset of audio events. They are therefore opposed to \textit{strong labels} that do provide time boundaries. Weak labels are enough to build an audio tagging (AT) system but innovative methods are required to fulfill a SED task which consists in determining the temporal location of target audio events.

With the availability of weakly-labeled large datasets such as Audioset~\cite{audioset}, comprised of Terabytes of audio extracts from video along with audio tags produced by YouTube users, the design of new models and methods to infer the missing temporal boundaries of events has gained momentum recently. In the recent literature, attention mechanisms are a popular choice~\cite{xu2017large,chen2018class}. In~\cite{xu2017large}, for instance, a single DNN is trained on a weakly-labeled dataset to perform both AT and SED. The authors proposed an architecture where the fully-connected output layer is duplicated with neuron units utilizing a sigmoid function and counterparts utilizing a softmax activation function that points out to which acoustic frames the model should attend to make AT predictions.

A different approach is geared towards weakly-supervised training and the Multiple Instance Learning (MIL,~\cite{dietterich1997solving}) paradigm in particular, in which this work is based. MIL is adapted to learning scenarios where the training examples are ambiguous: they are arranged in \textit{sets} called \textit{bags} comprised of both positive and negative instances of a class of interest. This is the case most of the time in SED where an audio recording is analyzed at short duration frames. An audio event may occur in some but not all the acoustic frames of a recording labeled as positive for the corresponding audio event class. MIL has recently been used in SED~\cite{kumar2016audio,morfi2018data}. In our own recent work~\cite{cances2018}, we generalized to a multi-label classification task a MIL-inspired loss function proposed for binary classification in~\cite{morfi2018data}. We trained a recurrent neural network for the polyphonic SED task of the DCASE 2018 challenge~\cite{dcase2018web}. While yielding interesting results, we noticed that some classes were not distinguished by this approach. This happens when two classes often occur simultaneously in the training subset, impeding the model to distinguish between them. To tackle this issue, we propose to introduce in the learning objective function a similarity penalty between the model predictions of the positive classes. We show in this article that a cosine-similarity penalty forces the model to output distinct class-specific temporal predictions. We apply this approach on the DCASE 2018 challenge data (task 4) and show that a penalized MIL objective brings a significant performance gain and outperforms an attention-based model of the same size.





This article is organized as follows. In Section~\ref{sec:mil}, we describe the MIL framework and derived existing methods applied to SED. We then introduce our contribution that consists in adding a cosine similarity penalty to the MIL loss function. We report our experiments in Section~\ref{sec:exp}, analyze the results and discuss limitations.

\section{Multiple Instance Learning}
\label{sec:mil}

The Multiple Instance Learning or Multi-Instance Learning (MIL) paradigm was first coined by Dietterich et al.~\cite{dietterich1997solving} for drug activity prediction. Their objective was to tackle what they called the "Multiple Instance Problem", in which the training examples are ambiguous: they are arranged in \textit{sets} called \textit{bags} comprised of both positive and negative instances of a class of interest. 
Since then, MIL has been studied, adapted, and applied to many tasks involving weakly labeled data. In~\cite{amores2013multiple}, a survey dedicated to multiple-instance (binary) classification, the author illustrates MIL with a clear example: image classification. If the target class to be detected is "beach", then a positive image is one where both sand and the sea appear in at least one region of the image. The other regions of the image, such as the sky, mountains or trees are, thus, negative instances.

In our case, weakly-supervised sound event detection, we face exactly the same MI situation. Indeed, instead of image regions as instances, we work with acoustic frames as instances. In a given recording weakly labeled as \textit{Dog}, some acoustic frames will comprise dog barking and others will not. Our end goal is to make predictions at frame-level, the so-called "strong" annotations, whereas the 
reference tags available for training are "weak labels" at file-level. 

More formally, we may denote a recording $X$ as a bag comprised of $T$ frames/instances: $X = \{\mathbf{x_1}, \ldots \mathbf{x_{T}}\}$, where the  $\mathbf{x_i}$ are feature vectors such as filter-bank coefficients, for instance. We are provided with a weakly-labeled training set of $M$ bags together with their global labels $y$: $\mathcal{T} = \{(X^1, y^1), \ldots, (X^M, y^M)\}$. In the case of binary classification, we have $y^î \in \{0, 1\}$, whereas in our case, we are concerned with multi-label classification where the $y$ labels are binary vectors of size the number of target classes $C$: $y \in \{0, 1\}^C$. Our goal is to estimate two classification functions: one to infer weak labels $\hat{y}$ and one to infer strong labels $\hat{y_t}$ at frame-level, where $t$ is a frame index. For the first one, the training set corresponds to the normal supervised learning scenario, and a binary cross-entropy ($\mathrm{binCE}$) loss can be used, for instance, to train a neural network to fulfill the task. 

For the second objective, the MIL one, a straightforward solution is the so-called "False Strong Labeling" (FSL hereafter), in which one considers that all the frames/instances are positive when a bag is labeled as positive: $y=1 \Rightarrow \forall t, y_t=1$. We will use FSL as our baseline method.
Of course, this strong approximation is suboptimal and MIL instead consists in considering that at least one instance is positive. Therefore, the highest scored instance in a bag should match the weak label of its bag, either a negative or a positive label.

This idea may be formalized in the following form for a given training example: $\max_{t}\hat{y}_{tc} = y_{c}$, where $\hat{y}_{tc}$ is the prediction score for frame $t$ and class $c$, and $y_{c}$ the reference tag for class $c$, either 0 or 1.

In order to train a neural network, we will then use a multi-label binary cross-entropy loss between $y_{c}$ and $\max_{t}\mathbf{\hat{y}_{tc}}$, as formulated in~\eqref{eq:maxloss} for a given training bag $k$:


\begin{equation}
\begin{gathered}
\mathrm{loss}(\{X^k, y_{c}^k\}) = \mathrm{binCE}(y_{c}^k, \max_{t}\mathbf{\hat{y}_{tc}^k}) 
\label{eq:maxloss}
\end{gathered}
\end{equation}

This loss, that we will refer to as the MIL loss henceforth, will be our baseline objective function. 

Recently, in~\cite{morfi2018data}, the authors proposed a variant of this loss function that they call a MIL MMM loss, where MMM stands for Max, Mean and Min to denote the three terms that compose their loss function. They noticed that the training of model with a MIL objective function focuses only on the highest scored frames and ignore the other ones. They proposed to add to the "max"-based term in the MIL loss formulation two extra terms: one that  takes into account the frames with the lowest prediction scores ("min"), which should tend towards zero, and another term ("mean") to impose a 0.5 average value of the model predictions throughout a recording. This last term is based on the na\"ive assumption that, in general, a specific event will be present in half of the frames. 

In one of our own submissions to the DCASE 2018 task 4, we successfully generalized this MMM loss to the multi-label SED task with a $C>2$ number of target classes~\cite{cances2018}. In the present work, we instead use the basic max MIL loss as formulated in~\eqref{eq:maxloss} as our baseline objective function. This is for several reasons, we wanted: 1) to keep the loss function as simple as possible, 2) to avoid potential unwanted interactions between the different loss terms and/or the penalty that we will introduce in the next section.




\begin{figure*}[htbp]
\centerline{\includegraphics[scale=0.35]{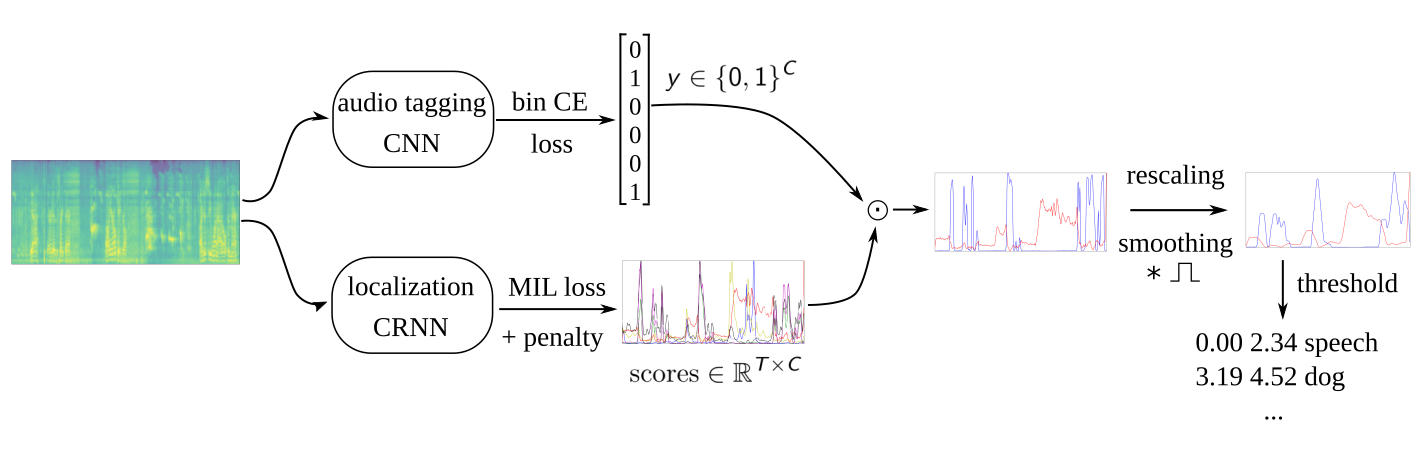}}
\caption{Overview of the proposed system. $C$: number of target classes, $T$: number of acoustic frames.}
\label{fig:system}
\end{figure*}

\section{Cosine-similarity as a penalty}

As we shall see in the experiment section, there is an issue with the MIL approach as formulated by the max MIL loss function in~\eqref{eq:maxloss}. This issue is specifically related to multi-label classification problems, where multiple classes occur in a given bag. In this type of problems, if two classes almost always co-occur in the training samples, there is no way for the classification model to distinguish between them, and the prediction scores for these two classes will be identical, or rather, highly correlated. In the DCASE 2018 task 4 training dataset in particular, the \textit{Dishes} and \textit{Frying} events co-occur in most of the training samples: on the 170 examples of \textit{Frying}, only 30 examples feature \textit{Frying} alone, all the others feature either \textit{Dishes}, \textit{Speech} or both. As we will report hereafter, the predictions for \textit{Dishes} and \textit{Frying} are highly correlated, although these two sounds are very distinct in nature: \textit{Dishes} corresponds to very short percussive sounds between dishes and cutlery whereas  \textit{Frying} are noisy long sounds that last the whole 10-s recordings. As a result, very poor accuracy was found for \textit{Dishes} confused with \textit{Frying}. This did not happen with \textit{Speech} since there are many more training examples for this class, and although speech never occur alone, speech co-occur with all the other target classes, guarantying enough variability to properly model \textit{Speech}.

In order to remedy this issue, we propose to introduce an additional term in~\eqref{eq:maxloss} to penalize the similarity between the frame-level prediction scores of the different target classes. This penalty takes the form of the cosine similarity measure between the scores predicted for the positive classes only. We do not penalize the similarity the negative classes' predictions since they all are expected to show low values constant in time, and, thus, are expected to be very correlated. We also chose to penalize positive similarity values only. The new loss function definition is given in~\eqref{eq:maxlosscos} for a given class $c=1,2, \ldots, C$ and a given bag for which we omit the $k$ index for ease of reading:


\begin{eqnarray}
\label{eq:maxlosscos}
\mathrm{loss}(\{X, y_{c}\})&{=}&\mathrm{binCE}(y_{c}, \max_{t}\mathbf{\hat{y}_{tc}}) \nonumber\\
&&{+}\:\alpha \, y_{c} \displaystyle \sum_{l \neq c}^C y_{l} \max(0, \cos(\mathbf{\hat{y}_{:l}},\mathbf{\hat{y}_{:c}}))
\end{eqnarray}
where $\alpha$ is a regularization weight, $y_{c}$ and $y_{l}$ the weak groundtruth binary labels for classes $c$ and $l$, and  $\mathbf{\hat{y}_{:l}}$, $\mathbf{\hat{y}_{:c}}$ the probability vectors for classes $c$ and $l$, where "$:$" denotes the frame varying index. Indeed, the cosine similarity is computed by taking the dot product along the time axis: 

\begin{eqnarray}
\cos(\mathbf{\hat{y}_{:l}},\mathbf{\hat{y}_{:c}}) = \frac{\mathbf{\hat{y}_{:l}}^T \cdot \mathbf{\hat{y}_{:c}}}{\norm{\mathbf{\hat{y}_{:l}}}\norm{\mathbf{\hat{y}_{:c}}}}
\end{eqnarray}




\section{System description and evaluation}
\label{sec:system}

Figure~\ref{fig:system} shows our proposed approach. It is based on two convolutional (recurrent) neural networks (CNN/CRNN): a CNN for audio tagging (top branch in the figure), trained with the binary cross-entropy loss, a CRNN for SED (bottom branch in the figure), trained with our custom MIL loss with and without the cosine penalty. The audio tagging network outputs a vector of binary predictions $\hat{y} \in \{0, 1\}^C$ whereas the SED one outputs a matrix of scores $S$ of size $T \times C$, where $T$ and $C$ are the number of acoustic frames and the number of classes, respectively. The binary predictions $\hat{y}$ are obtained using class-dependent thresholds that have been optimized with a genetic algorithm inspired by simulated annealing \cite{simulated_annealing}. This method reaches optimal values much faster than grid search. 

The two networks share a similar architecture, which will be described in details in Section~\ref{sec:exp}. The first part of the networks, the feature representation block, has the same architecture: repeated blocks comprised of convolution, batch-normalization, activation function (REctifier Linear Unit, ReLU), pooling and dropout layers. The main difference lies in the head of the networks. The AT network uses fully-connected decision layers with sigmoid as activation function of the output layer's units, as is standard for multi-label classification. For SED, we instead use a bi-directional Gated Recurrent Unit (GRU) layer followed by a "time-distributed" dense layer comprised of $C$ units, in order to obtain temporal predictions in the form of the score matrix $S$.

We tested an alternative regarding the representation extraction part of the SED network by replacing the ReLU activation by a trainable gated linear unit (GLU) that has been first introduced for language modeling~\cite{dauphin2016language} and successfully used for AT and SED~\cite{xu2017large}. GLU introduces an attention mechanism to all the convolution layers by controlling the amount of information of the data flow between layers. The input time-frequency representation or the intermediate feature maps from a given convolution layer are passed through two convolution layers in parallel, one with a linear activation and one with a sigmoid activation. They are then element-wise multiplied to each other. The sigmoid branch acts as a forget gate and allows the network to learn on what time-frequency sub-regions to attend to fulfill the task at best. It is worth noting that replacing ReLU by GLU doubles the size of the model in terms of number of trainable parameters.

Once audio tags and temporal predictions are obtained, we only keep the temporal predictions of the classes detected as positive by the audio tagging network. Then, they are rescaled to the $[0, 1]$ interval and smoothed with a sliding-average window. The final segments are obtained by using a single threshold for all the classes to detect the onsets and offsets of the events. We tested more complicated methods to derive these event segments such as hysteresis thresholds, which lead to slightly better results, but these methods require some tuning and we preferred to stick to a very simple threshold method for fair comparison between our approaches.

Since this work focuses on the event localization part of the task and not much on audio tagging, we used the same audio tag predictions in all our tests, for the sake of fair comparison between the SED models. We ran our best audio tagging model once and used its tag predictions for all our localization models.

Besides comparing the performance of SED models trained with and without the cosine penalty, we also trained a baseline model with "False Strong Labels" (FSL), a model with the same architecture as the SED CRNN. As written earlier, this baseline consists in considering that the strong labels correspond to the weak labels, meaning that if there is the \emph{Dog} tag for a file, then all the acoustic frames of this file are considered positive in regards to the \emph{Dog} class.

We also tested a model similar to the DCASE 2017 winning solution based on GLUs~\cite{xu2017large}. In this approach, that we will refer to as the ATT model (ATT for attention), a single DNN is trained on a weakly-labeled dataset to perform both AT and SED. The authors proposed an architecture where the fully-connected output layer is duplicated with neuron units utilizing a sigmoid function and counterparts utilizing a softmax activation function that points out to which acoustic frames the model should attend to make AT predictions. Sigmoid allows for multi-label sound classification and softmax is responsible for pointing out to which acoustic frames the model should attend to make AT predictions. The softmax layer, thus, performs event localization. 

Finally, performance was assessed in terms of an event-based metric with a 200~ms collar on onsets and a 200~ms/20\% of the events length collar on offsets. This metric is a F-measure, called F-score hereafter, that was proposed and used in the framework of the DCASE task 4 challenge~\cite{Serizel2018}.




\section{Experimental setup}
\label{sec:exp}

\subsection{Audio material}

In this work, we used the data from the DCASE 2018 Task 4 challenge~\cite{Serizel2018} and more precisely the labeled in-domain set. The corpus is comprised of 10-second clips extracted from Youtube user videos and are part of the Audioset corpus~\cite{audioset}. The recordings most often contain several overlapping event categories. The audio events correspond to a set of 10 sound categories occurring in domestic environments: Speech, Dog, Cat, Alarm/Bell ringing, Dishes, Frying, Blender, Running water, Vacuum cleaner, and Electric shaver/toothbrush. 

We use the weakly-labeled training subset of 1578 clips (2244 class occurrences, 4.4h of audio data), for which the weak annotations (audio tags) have been verified and cross-checked manually, and the test subset comprised of 288 files for which strong labels are available. We will report performance results on this subset, which is different from the challenge evaluation set that is not publicly available.

We used 20\% of the training set as a development set in order to set hyperparameters such as the number of epochs, the learning rate, etc. 


\subsection{Audio features}


As input to the networks, 64 log-Mel filter-bank (F-BANK) coefficients were extracted every 23~ms on 100~ms duration frames, with 20~Hz and 11025~Hz as minimum and maximum frequency values to compute the Mel bands, respectively. Hence, for each 10-second file, a $431 \times 64$ matrix is extracted. This matrix is used as a single input image fed to the networks.

Different normalization and feature scaling methods were tested as pre-processing such as \textit{global mean removal}, \textit{mean and variance standardization}, but no gains were observed compared to using raw F-BANK.

\begin{figure}[htbp]
\centerline{\includegraphics[scale=0.7]{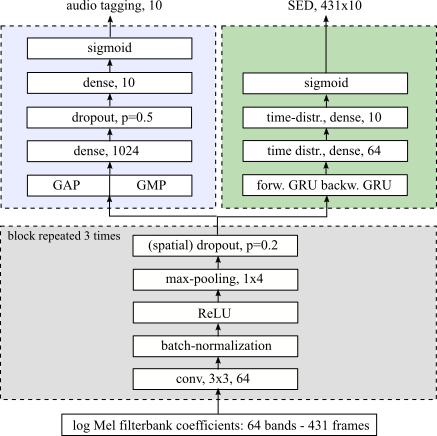}}
\caption{Architecture of the audio tagging and SED Networks.}
\label{fig:networks}
\end{figure}

\begin{center}
\begin{table*}[ht]
\caption{Global macro and class-wise SED F-scores on the Test subset. FSL: False Strong Labeling, ATT: attention-based AT and SED model. \emph{+cos}: GLU-MIL trained with the cosine penalty, \emph{+AT oracle}: GLU-MIL+cos when using the ground truth audio tags instead of predicted ones. Std: standard deviation. 1-best: best model.} 
\label{tab:results}
\begin{center}
\begin{tabular}{lcccccccc}
\toprule
    Approach & FSL & ATT & MIL & GLU-MIL & +cos & +cos (1-best) & +AT oracle & +AT oracle (1-best)\\
    \midrule
    F-score (\%) & 15.27 & 16.80 & 28.90 & 30.15 & 31.50 & 34.75 & 37.93 & 42.39 \\
    Std (F-score) (\%) & \_ & \_ & \hphantom{1}1.95 & \hphantom{1}0.79 & \hphantom{1}1.83 & \_ & \hphantom{1}2.36 & \_ \\ 
    \midrule
    Alarm / bell / ringing & \hphantom{1}2.5 & 15.8 & 32.9 & 34.6 & 32.8 & 38.4 & 36.2 & 43.3\\
    Blender & 10.7 & 14.0 & 20.6 & 20.7 & 27.4 & 26.4 & 49.7 & 49.5  \\
    Cat & \hphantom{1}3.3 & \hphantom{1}3.2 & 49.3 & 45.6 & 47.3 & 53.2 & 47.3 &  54.4 \\
    Dishes &  \hphantom{1}0.0 & 18.0 & \hphantom{1}0.0 & \hphantom{1}0.0 & \hphantom{1}4.1 & 16.0 & \hphantom{1}6.1 & 25.1\\
    Dog & \hphantom{1}2.4 & \hphantom{1}9.9 & 16.4 & 27.4 & 28.6 & 26.7 & 31.8 & 29.4\\
    Electric shaver / toothbrush & 40.9 & 13.5 & 27.8 & 30.9 & 37.8 & 41.7 & 37.7 & 41.9 \\
    Frying & 30.6 & 14.9 & 27.8 & 34.1 & 27.0 & 31.2 & 40.8 & 45.9 \\
    Running water & \hphantom{1}8.2 & 14.7 & 14.9 & 13.1 & 13.7 & 15.4 & 20.9 & 22.5 \\
    Speech & \hphantom{1}0.0  & 36.0 & 36.6 & 35.8 & 37.0 & 38.4 & 36.8 & 38.4 \\
    Vacuum cleaner & 54.3 & 28.0 & 62.7 & 59.2 & 59.3 & 60.3 & 72.0 & 73.4 \\
\bottomrule
\end{tabular}
\end{center}
\end{table*}
\end{center}

\begin{center}
\begin{table*}[ht]
\caption{Results on the Eval subset of the models that performed best on the Test subset. The two first ranked teams' performance are shown in the last two columns.} 
\label{tab:eval}
\begin{center}
\begin{tabular}{lcccccc}
\toprule
    Approach & Official Baseline & MIL & GLU-MIL & GLU-MIL+cos & JiaKai~\cite{Lu2018} & Liu~\cite{Liu2018}\\
    \midrule
    F-score (\%) & 10.8 & 18.86 &  22.60 & 26.20 & 32.4 & 29.9 \\
    \midrule
    Alarm / bell / ringing & \hphantom{1}4.8 & 30.1 & 29.0 & 30.4 & 49.9 & 46.0 \\
    Blender & 12.7 & 28.8 & 23.3 & 27.7 & 38.2 & 27.1 \\
    Cat & \hphantom{1}2.9 & 22.8 & 28.7 & 30.3 & \hphantom{1}3.6 & 20.3 \\
    Dishes & \hphantom{1}0.4 & \hphantom{1}1.0 & \hphantom{1}0.0 & 19.0 & \hphantom{1}3.2 & 13.0 \\
    Dog & \hphantom{1}2.4 & 20.1 & 19.8 & 20.9 & 18.1 & 26.5 \\
    Electric shaver / toothbrush & 20.0 & \hphantom{1}7.7 & \hphantom{1}6.2 & 19.1 & 48.7 & 37.6 \\
    Frying & 24.5 & \hphantom{1}0.0 & 27.6 & 21.2 & 35.4 & 10.9 \\
    Running water & 10.1 & 17.9 & 13.3 & 13.2 & 31.2 & 23.9 \\
    Speech & \hphantom{1}0.1 & 36.7 & 37.6 & 35.0 & 46.8 & 43.1 \\
    Vacuum cleaner & 30.2 & 23.5 & 40.6 & 45.2 & 48.3 & 50.0 \\
\bottomrule
\end{tabular}
\end{center}
\end{table*}
\end{center}

\subsection{Neural networks and post-processing}


As stated in Section~\ref{sec:system}, we used two networks, one for AT, one for SED. They architecture is shown in Figure~\ref{fig:networks}.

The two networks share a similar feature representation extraction part: three convolution blocks, each comprised of 64 filters (3x3) followed by batch-normalization, a Rectifier Linear Unit (ReLU) activation function and max-pooling (4) along the frequency axis only. 

In the case of the AT model, we used 2-d spatial dropout (\textit{p}=20\%), which means that a percentage of entire feature maps are randomly set to zero instead of isolated pixels as in the case of standard dropout. Spatial dropout is recommended when the input data presents high local correlations. Our preliminary AT experiments tended to show slight improvements thank to spatial dropout but this would need to be further explored and this is not the focus of the present work. No dropout was used in the convolution blocks of the SED model as we found slight decrease in performance when using it. To prevent from overfitting, we  train the SED model for ten epochs only. In any case, the loss function curves show a plateau after around seven epochs.

For the AT model, the last convolution block is followed by 2-d average- and 2-d max-global pooling layers (GAP and GMP), then by a dense one with 1024 units, dropout (\textit{p}=20\%) and a 10-unit dense output layer with sigmoid activation. In our preliminary tests, this network was found to perform slightly better than a model using a recurrent layer. It yielded 85.84\% and 82.86\% f1-scores on our training and validation subset (proportion: 80/20 \% of the weakly labeled training set).

For the SED model, the feature representation block is followed by one bi-directional recurrent layer in the form of 64 Gated Recurrent Units GRU with 10\% dropout on the input and on the recurrent states, a tanh activation function, then a time-distributed dense layer of 64 units with 10\% dropout on the input, followed by a second and final output dense layer comprised of 10 units with sigmoid activation. This architecture outputs temporal predictions in the form of a matrix of dimension $431 \times 10$ as described earlier. The number of trainable parameters is about 100k weights. When using GLUs, this number doubles to 200k weights.

All the models are trained on weak labels for 100 epochs for the AT, FSL and ATT models with early stopping after 15 epochs patience with a minimum loss difference threshold of $1\mathrm{e}{-4}$. The MIL-based models were trained for ten epochs only. We used the Adam optimizer with default parameters. 

Regarding the cosine penalty, we will report results obtained with an $\alpha=0.1$ regularization weight. This value was found by carrying out a coarse grid search.

The score curves are individually rescaled to $[0, 1]$. The final event segments are obtained by first smoothing the score curves with a moving-average filter of size 19 frames. Second, the curves are binarized with a 0.03 threshold. Neighbor segments are merged when separated by less than 200 ms, which is the tolerance margin used for evaluation. Source code in Keras with Tensorflow as backend is available online\footnote{https://github.com/topel/ijcnn19\_submission}.

\section{Results}


\subsection{Results on the test subset}

The audio tagging CNN achieved 77.2\% AT macro F-score on the test subset. The best recognized class is \emph{Speech} (91\% F-score) and the least well recognized ones are \emph{Dishes} and \emph{Blender} (60\% F-score). In the case of \emph{Dishes}, recall is low (53\%) indicating that the model misses occurrences, whereas for \emph{Blender}, recall and precision scores are similar. 

As described here-above, for the SED experiments reported here-after, we only consider the temporal predictions of the classes detected by the AT network. The impact of AT errors on SED performance will be assessed by using the ground truth weak labels of the test subset (\textit{oracle} system) instead of the error-prone AT predictions.

Table~\ref{tab:results} shows the localization performance results on the test subset in terms of F-score for the FSL baseline, ATT and our MIL-based approaches. For each of the latter, we trained ten models and show the averaged global F-scores with standard deviation and the averaged class-wise F-scores. We also show the F-scores obtained by the best model (1-best) out of the ten ones trained with the penalty (+cos). 

FSL achieved a 15.27\% F-score and class-wise results similar to the FSL baseline provided by the challenge organizers that yielded 14.06\% F-score~\cite{Serizel2018}. By observing the duration distributions for each class~\cite{Serizel2018}, it appears that, as expected, FSL achieves very poor results for short duration events, namely \emph{Alarm/bell/ringing}, \emph{Cat}, \emph{Dishes}, \emph{Dog}, \emph{Speech}, and better results for long events that last the whole 10-s recordings, namely \emph{Blender}, \emph{Electric shaver/toothbrush}, \emph{Frying}, \emph{Vacuum cleaner}, and to a less extent \emph{Running water}, which has a more uniform duration distribution. This is of course due to over-generalization since the FSL model is trained with weak labels converted as strong labels and, thus, many negative frames are wrongly considered as positive for short events.



\begin{figure*}[htbp]
\centerline{\includegraphics[scale=0.37]{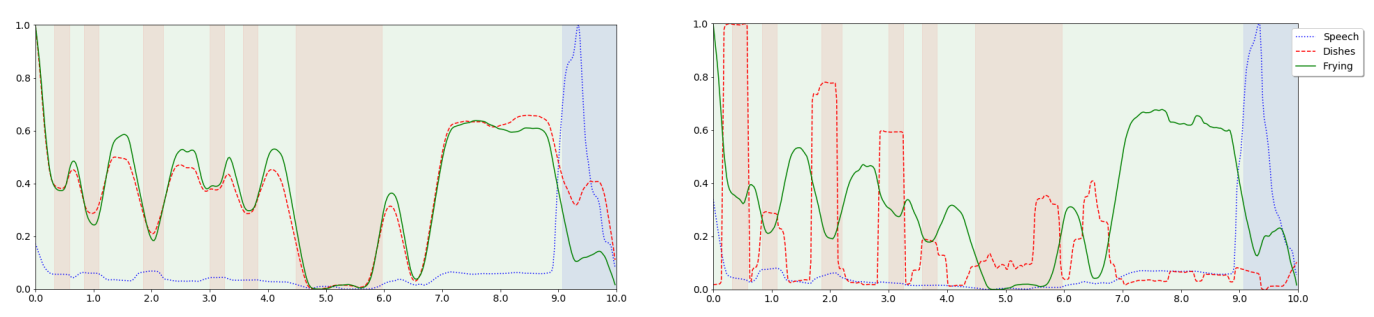}}
\caption{Prediction curves of the true positive classes: \emph{Speech} (in blue, dotted line), \emph{Dishes} (in red, dashed line) and \emph{Frying} (in green, plain line), for the Y5J603SAj7QM\_210.000\_220.000 test file. Left: when using MIL without penalty, Right: with penalty. The groundtruth is indicated as segments coloured with the same class colours as the prediction curves. The true \emph{Frying} segment, in light green, lasts the whole 10 seconds.}
\label{fig:Y5}
\end{figure*}

\begin{figure*}[htbp]
\centerline{\includegraphics[scale=0.37]{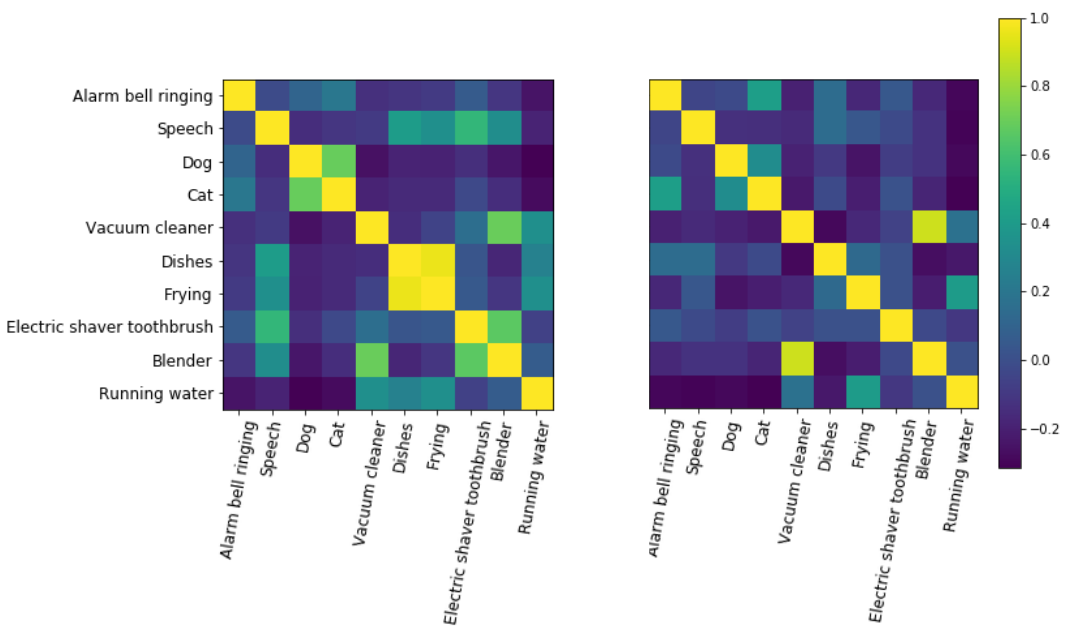}}
\caption{Pearson correlation coefficients between the ten classes predictions computed on the test subset, obtained by two models trained with the MIL loss. Left: without penalty, Right: with penalty. The correlation between \emph{Dishes} and \emph{Frying} is greatly reduced when using the penalty: from 96\% to 14\%.}
\label{fig:corr}
\end{figure*}

Basic MIL models (MIL column in Table~\ref{tab:results}) outperformed the FSL baseline by a large margin mainly by achieving better results on short duration events. For example, FSL and MIL yielded 3.3\% and 49.3\% F-scores for \emph{Cat}, respectively. MIL also outperformed the attention model ATT that yielded a 16.80\% F-score only. It should be noted that we used the same audio tag predictions as with the other models for the sake of fair comparison. As a side information, the attention model also led to poor performance in AT with 68.2\% F-score. This AT score indicates that strong overfitting occurred. These low performance in both AT and SED are surprising given that the authors reached much better figures with this kind of approach in the 2017 DCASE edition on the same task~\cite{xu2017large}. This could be explained by the fact that here we used a much smaller model with a number of parameters comparable to the GLU-MIL models for the sake of fair comparison. Other differences with their work are feature standardization on each frequency bin and mini-batch data balancing that are not used in our work. It is notable, though, that ATT reached a 18\% F-score for \emph{Dishes}.


The GLU-MIL column in the table gives the averaged performance of  ten CRNNs where GLU replaces ReLU after all the convolution layers. As can be seen, GLUs further improved performance of MIL by reaching 30.15\%$\pm0.79$\% global F-score. This confirms the interesting impact of introducing an attention mechanism in the network architecture. Nevertheless, \emph{Dishes} is not detected at all. Adding the cos penalty to GLU-MIL did not significantly improve overall performance but allowed \emph{Dishes'} score to rise from zero achieving a 4.1\% value and even a 16\% with the best model out of ten (\emph{1-best}). The latter achieved a 34.75\% F-score. By way of comparison, the system ranked first in the competition achieved a score of 25.9\% on the same subset~\cite{Lu2018}. Not all the \emph{+cos} models had a positive score for \emph{Dishes} showing that training with the penalty does not always ensure the removal of the \emph{Dishes/Frying} confusion. There is room for improvement regarding this point. The second last column shows the averaged results of the same GLU-MIL+cos models but when AT predictions are replaced by the ground truth tags: GLU-MIL+cos+oracle. In this case, errors are only due to the SED models. F-score increased by about 20\% relative compared to GLU-MIL+cos and reached a 37.93\%$\pm2.36$\% topline value showing that AT errors have a great impact on the final SED performance. Finally, the last column shows the \emph{1-best} model performance and a 42.39\% topline value.

\subsection{The cosine penalty tends to decorrelate the predictions}

It is remarkable that FSL, MIL and GLU-MIL completely fail for \emph{Dishes}. By observing the localization predictions, it appears that \emph{Dishes} and \emph{Frying} are not distinguished. In the training subset, about 46\% of the \emph{Dishes} samples also contain \emph{Frying}. There are even more files featuring \emph{Dishes} and \emph{Speech}, namely 52\%, and 32\% with the three classes together. \emph{Dishes} is not confused with \emph{Speech} probably because there are many more \emph{Speech} files than \emph{Dishes} files: 550 versus 184 files.

To illustrate the \emph{Dishes/Frying} confusion, we plotted an example in Figure~\ref{fig:Y5}. In the left-hand side plot, we plotted the smoothed and rescaled probability curves obtained with GLU-MIL for a test file example (id: Y5J603SAj7QM\_210.000\_220.000). Three curves are shown, for \emph{Speech} (in blue, dotted line), \emph{Dishes} (in red, dashed line) and \emph{Frying} (in green, plain line). We also represent the groundtruth segments with transparent rectangles of the same colours as the curves. As expected, in the left figure, \emph{Dishes'} and \emph{Frying}'s curves are almost identical and they share a very high Pearson correlation coefficient of 87\%. 

The right-hand side figure shows the same curves when training a GLU-CRNN with the penalty: GLU-MIL+cos. This time the two curves are completely different with a negative (-20\%) correlation value. Clear peaks match the true \emph{Dishes} segments. The cosine similarity penalty played its decorrelating role and improved the detection of \emph{Dishes}.




To measure this effect on all the classes, Figure~\ref{fig:corr} illustrates the Pearson correlation coefficients between the predictions of the ten classes computed on the test subset, obtained by two models trained with the MIL-max loss: without penalty (left-hand side), with penalty (right-hand side). Without penalty, some classes are pretty correlated, such as \emph{Dog} and \emph{Cat}, \emph{Blender} and \emph{Vacuum cleaner}. A 96\% correlation appears for \emph{Dishes} and \emph{Frying} meaning that the two classes are almost undistinguished. This correlation is greatly reduced to a 14\% value as can be seen in the right-hand side figure. Overall, when averaging all the positive correlations of these matrices, the global correlation value reduces from 12.6\% to 6.2\% between the two models. 

\subsection{Results on the Eval subset}

Table~\ref{tab:eval} shows the results on the official evaluation subset of DCASE 2018 obtained with our systems MIL, GLU-MIL and GLU-MIL+cos, together with the ones of the two best ranked systems: JiaKai~\cite{Lu2018} and Liu \textit{et al}~\cite{Liu2018}. For our models, we report performance of single models, namely the ones that performed the best on the test subset. The GLU-MIL+cos approach outperforms our other MIL models with a 26.20\% F-score. This system would rank third in the competition\footnote{http://dcase.community/challenge2018/} (team ranking) as the first two systems reached 32.4\%~\cite{Liu2018} and 29.9\%~\cite{Lu2018} F-scores. The first one used a semi-supervised approach where the unlabeled in-domain data were automatically labeled and added to the training subset. This system is a CRNN based on a mean-teacher learning procedure. In the present work, we did not explore the use of the unlabeled data, this remains for future work. Regarding the system ranked second, it consists of an ensemble of ten models trained in a strongly-supervised way: a threshold-based event activity detector was used to strongly annotate the weakly-annotated subset.   

\section{Discussion}

With these results and observations, we showed that MIL is an effective paradigm to handle multi-label SED in a weakly-supervised setting and that the cosine penalty increases the discriminative power of the network. 

Penalizing the inner product between the score vectors of the positive classes tends to force events to not overlap. This effect is visible in the right-hand side plot in Fig.~\ref{fig:Y5}: the \emph{Dishes} peaks (red) are in phase with sharp decreases of the \emph{Frying} scores (green). In the \emph{Speech} segment also (in blue), the \emph{Frying} and  \emph{Dishes} curves decreased to zero and only the \emph{Speech} blue curve has large values although the model should also predict \emph{Frying} in that segment. This is a drawback of using this penalty in cases where different events overlap. 

However, in our experiments on this specific dataset, this effect is globally positive. The accumulated duration of overlapping events in the test subset totals 3.81~min whereas it reaches 11.82~min in the GLU-MIL predicted segments. Using the penalty, this value significantly decreases to 7.66~min and even 5.51~min with +oracle, a value close to 3.81~min.

Instead of penalizing colinear predictions, a possibility would be to penalize the similarity between the prediction distributions, for example with the Kullback-Leibler divergence. 


    
    


\section{Conclusion}

In this paper, we proposed to tackle weakly-supervised SED through the Multiple Instance Learning paradigm. Although effective, an issue was observed due to the use of MIL in a multi-label context: the model is not able to distinguish between sound classes that most often occur simultaneously in the training subset.  We  enhanced a baseline MIL loss function by introducing a cosine similarity penalty between the score outputs of a convolutional recurrent neural network. 

Experiments conducted on the data set of Task 4 of the DCASE 2018 challenge led to the following conclusions: 

\begin{itemize}
    \item MIL is an effective paradigm to handle multi-label SED in a weakly-supervised setting with very different types of sounds (short and long sounds, noisy and harmonic/voiced sounds), 
    \item the cosine penalty significantly increases the discriminative power of the network by decorrelating the temporal predictions of the different classes. It seems to lead to a more balanced performance of the individual classes.
\end{itemize}

Our experiments also confirm recent results in the literature that the introduction of an attention mechanism in the form of Gated Linear Units allows significant performance gains.

    
    
        
Limitations were described. In particular, although it has a global positive impact on the results, we observed that penalizing the inner product of predictions tends to ensure that events do not overlap. This is a drawback for SED since events often co-occur, in particular in the case of a continuous sound such as \emph{Running water} or \emph{Frying}, other sound events can occur simultaneously. A future avenue for research will be to explore penalties on prediction distributions rather than directly on predictions. The Kullback-Leibler divergence, for example, could be a good candidate for such a penalty. Another improvement in our method would be to achieve the same performance but using a single neural network. Finally, another research direction is the use of semi-supervised approaches to turn the MIL approach more robust to unseen recordings. This could take the form of  a mean-teacher learning strategy as the one used by the first ranked team, for instance.

\section*{Acknowledgment}

This work was partly supported by the LABEX AMIES-UMS 3458, within the PEPS REP4SPEECH project, and the Agence Nationale de la Recherche LUDAU (Lightly-supervised and Unsupervised Discovery of Audio Units using Deep Learning) project (ANR-18-CE23-0005-01).  







\bibliographystyle{IEEEtran}
\bibliography{biblio_sed}

\end{document}